\documentclass{article}
\usepackage[margin=1in]{geometry}
\usepackage[onehalfspacing]{setspace}
\usepackage{amsmath,mathtools,amssymb}
\usepackage{pgfplots}
\usepackage{subcaption}
\usepackage{booktabs}
\usepackage[braket, qm]{qcircuit}
\usepackage{graphicx}
\usepackage{authblk}
\usepackage{xcolor}
\usepackage{amsthm}
\usepackage{float}

\newtheorem{remark}{Remark}

\DeclarePairedDelimiter\ceil{\lceil}{\rceil}

\title{Efficient Implementation of a Quantum Search Algorithm for Arbitrary \( N \)}

	\author[1]{Alok Shukla \thanks{Corresponding author.}}
	\author[2]{Prakash Vedula}
	\affil[1]{School of Arts and Sciences, Ahmedabad University, India}
	\affil[1]{alok.shukla@ahduni.edu.in}
	\affil[2]{School of Aerospace and Mechanical Engineering, University of Oklahoma, USA}
	\affil[2]{pvedula@ou.edu}
	
\pgfplotsset{compat=1.18}

\begin{document}
\maketitle

\begin{abstract}
This paper presents an enhancement to Grover's search algorithm for instances where the number of items (or the size of the search problem) $N$ is not a power of 2. By employing an efficient algorithm for the preparation of uniform quantum superposition states over a subset of the computational basis states, we demonstrate that a considerable reduction in the number of oracle calls (and Grover's iterations) can be achieved in many cases. For special cases (i.e., when $N$ is of the form such that it is slightly greater than an integer power of 2),  the reduction in the number of oracle calls (and Grover's iterations) asymptotically approaches 29.33\%. This improvement is significant compared to the traditional Grover's algorithm, which handles such cases by rounding $N$ up to the nearest power of 2. The key to this improvement is our algorithm for the preparation of uniform quantum superposition states over a subset of the computational basis states, which requires gate complexity and circuit depth of only $ O (\log_2 (N)) $, without using any ancilla qubits.

\end{abstract}

\section{Introduction}

One of the most well-known quantum algorithms is Grover's search algorithm \cite{grover1996fast, nielsen2002quantum}, which provides a quadratic speedup for unstructured search problems. Traditional implementation of Grover's search algorithm (which is a special case of the more general Amplitude Amplification algorithm \cite{Brassard_2002}) requires the number of items $N$ (i.e., the size of search space $N$) to be a power of $2$. 
The assumption that $N=2^n$ simplifies the creation of uniform superposition states (a key step in Grover's algorithm), as it allows for the straightforward application of $n$ Hadamard gates to the input state $\ket{0}^{\otimes n}$. 
When the size of the search space \( N \) is not a power of $2$ (say, $2^{n-1} < N < 2^n$), the traditional Grover's search algorithm relies upon considering an expanded search space such that the size of the search space rounds up to the nearest power of $2$, i.e., $2^n$. One reason for this was the lack of an efficient algorithm for the preparation of uniform superposition states when $N$ is not an integer power of $2$. This paper presents an improvement to the traditional implementation of Grover's search algorithm for cases where the size of the search space \( N \) is not a power of $2$. 
By employing a recently proposed efficient algorithm for the preparation of uniform quantum states over a subset of the computational basis states \cite{shukla2024efficient}, we demonstrate that in comparison to the traditional implementation of Grover's search algorithm,  considerable reduction in the number of oracle calls and Grover's iterations can be achieved in many cases when $N$ is not of the form of an integer power of $2$. Specifically for instances where 
$N$ is slightly greater than an integer power of $2$, the reduction in the number of oracle calls asymptotically approaches 29.33

The key to our improvement is a novel algorithm for the preparation of uniform quantum superposition states over a subset of the computational basis states. This algorithm achieves the desired state preparation with a gate complexity and circuit depth of only $O(\log_2 (N))$ and uses only $\ceil{\log_2 ( N))}$ qubits \cite{shukla2024efficient}.  Moreover, this algorithm does not need any ancilla qubits.  This efficiency in the preparation of uniform superposition states translates directly into the implementation of an efficient circuit for the modified Grover's algorithm with fewer oracle calls and Grover's iterations for several cases of $N$.

\section{Amplitude Amplification and Grover's Search Algorithms}

Following the reference \cite{Brassard_2002}, consider an $N$-dimensional Hilbert space $\mathcal{H}$ with orthonormal basis $\mathcal{B} = \{\ket{i}\}_{i=0}^{N-1}$, with $N=2^n$. Let $\mathcal{H}_g$ and $\mathcal{H}_b$ denote the set of ``good'' and ``bad'' states respectively.
Let $\mathcal{P}$ be a Hermitian projection operator on good states,  
\[
\mathcal{P} = \sum_{\ket{x} \in \mathcal{H}_g \cap \mathcal{B} } \, \ket{x}\bra{x}.
\]
Let $\mathcal{A}$ be a unitary operator such that $\mathcal{A} \ket{0} = \ket{\Psi}$.
The goal is to evolve an initial state $\ket{\psi} \in \mathcal{H}$ into $\mathcal{H}_g$.
Let
$\ket{\psi} =   \cos(\theta) \ket{\psi_b} + \sin(\theta) \ket{\psi_g},$
where $\theta = \arcsin(\|\mathcal{P} \ket{\psi}\|)$, $\ket{\psi_g} \in \mathcal{H}_g$, $\ket{\psi_b} \in \mathcal{H}_b$.
Define the unitary operator $Q = -U_{\psi} U_{\mathcal{P}}$, where
\begin{align} \label{eq:defUops}
U_{\psi} =  \mathcal{A} \, U_0 \, \mathcal{A^{\dagger}}  =  I - 2 \ket{\psi}\bra{\psi}, \quad U_0 = \left(I - 2 \ket{0}\bra{0}\right), \quad
U_{\mathcal{P}} = I - 2 \mathcal{P}.
\end{align}
It can be shown that the unitary operator $Q$ acts as a rotation by $2\theta$:
\[
Q = 
\begin{pmatrix}
\cos(2\theta) & -\sin(2\theta)\\
\sin(2\theta) & \cos(2\theta),
\end{pmatrix}
\]
in the subspace spanned by $\ket{\psi_b}$ and $\ket{\psi_g}$ (taking them as an ordered basis).
After $k$ applications of $Q$:
\[
Q^k \ket{\psi} = \cos((2k+1)\theta) \ket{\psi_b} + \sin((2k+1)\theta) \ket{\psi_g},
\]
the probability of measuring a good state is $\sin^2((2k+1)\theta)$, maximized by: $k  = \left\lfloor \frac{\pi}{4\theta} \right\rfloor = \left\lfloor \frac{\pi}{4} \sqrt{\frac{N}{M}}  \right\rfloor  $, where $M$ is the cardinality of ``good'' basis states, i.e., of the set $\mathcal{H}_g \cap \mathcal{B}$.

Thus, amplitude amplification process iteratively increases the amplitude of the good states. When, $\mathcal{A} = H^{\otimes n}$ and 
 $\mathcal{A} \ket{0} = \ket{\Psi} = \frac{1}{\sqrt{N}} \sum_{i=0}^{N-1} \ket{i}$, we get the standard Grover's quantum search algorithm for the search space of size $N= 2^n$.

\begin{figure}
    \centering
    \includegraphics{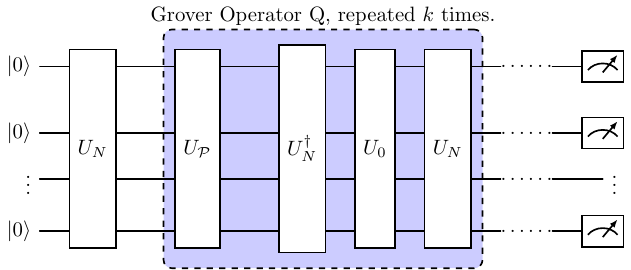}
    \caption{Schematic for an efficient implementation of the quantum search algorithm for arbitrary $N$. If $N$ is an integer power of $2$, then $U_N = H^{\otimes n}$ and it reduces to the traditional Grover's search algorithm. In the general case when $N$ is not an integer power of $2$, $U_N$ represents the unitary operator corresponding to the quantum circuit created by Algorithm 1  in \cite{shukla2024efficient} for preparation of uniform quantum states, i.e.,  $U_{N} \ket{0}^{\otimes n} = \frac{1}{\sqrt{N}}  \,  \sum_{j=0}^{N-1} \, \ket{j}$. Further, in this case, the quantum circuit for $U_N$ has gate complexity and circuit depth of order $O(\log_2 (N))$ and does not require any ancilla qubits. The unitary operators $U_{\mathcal{P}}$ and $U_0$ are defined in Eq.~\eqref{eq:defUops}.}

    \label{fig:schematic}
\end{figure}

\section{Efficient implementation of quantum search algorithm}
An efficient quantum algorithm for the preparation of uniform superposition state $\ket{\Psi} = \frac{1}{\sqrt{N}}\sum_{j = 0}^{N - 1} \ket{j}$,  with $ 2 < N < 2^n $ and $ N \neq 2^r $ for any $ r \in \mathbb{N} $, was recently provided \cite{shukla2024efficient}.
This algorithm prepares the desired quantum uniform superposition state with a gate complexity and circuit depth of only \( O(\log_2 (N)) \) and requires only \( \lceil \log_2 (N) \rceil \) qubits. In the following, we show that the traditional Grover's search algorithm can be improved by up to $29.3 \, \%$ using the approach for efficient preparation of uniform superposition states for arbitrary $N$ provided in \cite{shukla2024efficient}.

If $ N =  2^n$, then let $U_N = H^{\otimes n}$, otherwise if $N$ is not an integer power of $2$, then define $n = \ceil{\log_2 (N)}$ and 
let $U_{N}$ be the unitary operator corresponding to the quantum circuit created by Algorithm 1 in \cite{shukla2024efficient}, i.e.,  $U_{N} \ket{0}^{\otimes n} = \frac{1}{\sqrt{N}}  \,  \sum_{j=0}^{N-1} \, \ket{j}$.
Then, replacing $\mathcal{A}$ in the amplitude amplification algorithm by $U_N$ provides a more efficient method for search compared to the traditional Grover's search algorithm. A schematic for efficiently implementing the quantum search algorithm for an arbitrary 
$N$ is shown in Fig.~\ref{fig:schematic}.

As noted earlier, when \( N \) is not a power of 2, the traditional approach rounds \( N \) up to the nearest power of 2, \( 2^n \), where \( n = \lceil \log_2 (N) \rceil \), with the number of iterations for the old approach (i.e., traditional Grover's search) and the new approach (i.e., the proposed approach using Algorithm 1 in \cite{shukla2024efficient}) being \( T_{\text{old}} = \left\lfloor \frac{\pi}{4} \sqrt{\frac{2^n}{M}} \right\rfloor \) and \( T_{\text{new}} = \left\lfloor \frac{\pi}{4} \sqrt{\frac{N}{M}} \right\rfloor \), respectively.
An improvement factor $f$ can be defined by the  ratio of iterations as given by $$
f  =
\frac{T_{\text{old}}}{T_{\text{new}}} = \frac{\left\lfloor \frac{\pi}{4} \sqrt{\frac{2^n}{M}} \right\rfloor}{\left\lfloor \frac{\pi}{4} \sqrt{\frac{N}{M}} \right\rfloor}.
$$
 Clearly, asymptotically, as $\frac{N}{M}$ grows larger, one can obtain
\[
f  =
\frac{T_{\text{old}}}{T_{\text{new}}} \approx \sqrt{\frac{2^n}{N}}.
\]
Since \( 2^{n-1} < N \leq 2^n \),
\[
1 \leq f  = \sqrt{\frac{2^n}{N}} < \sqrt{2} \approx 1.414.
\]
The percentage improvement 
 $\eta$, which represents the percentage reduction in oracle calls and Grover iterations (in comparison to traditional implementation of Grover's search algorithm), is given by 
 $ 
 \eta = \left( 1 - \frac{1}{f}\right) \times 100 \%$. 
 It is easy to see that
 \[
0  \leq \eta < \left(1 - \frac{1}{\sqrt{2}}\right) \times 100\% \approx 29.3\%.
 \]
Clearly, when \( N \approx 2^{n-1} \) and as $N$ grows larger, the resulting improvement asymptotically approaches approximately 29.3\%.

\begin{figure}[ht]
    \centering
    \includegraphics[width=0.48\textwidth]{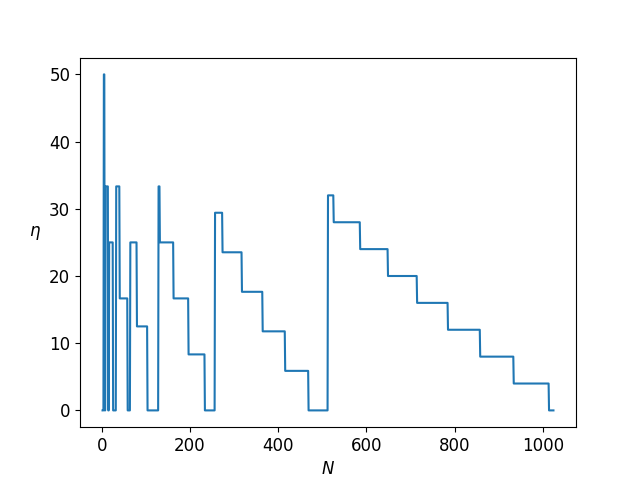}
    \includegraphics[width=0.48\textwidth]{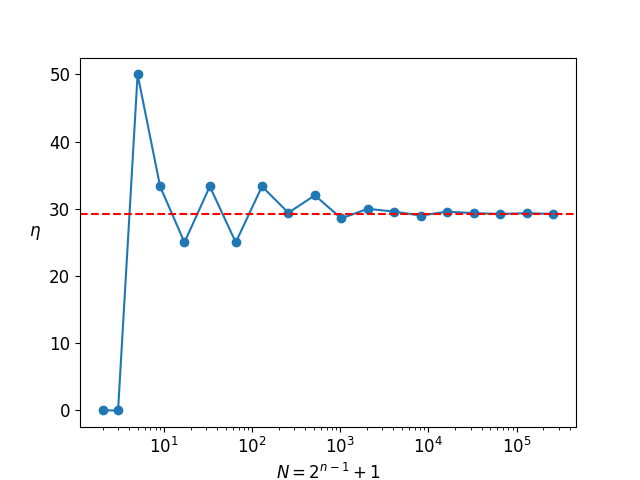}
    \caption{Percentage improvement, $\eta$, representing a percentage reduction in oracle calls and Grover iterations for $M=1$ and arbitrary $N$ (left) and for $N = 2^{n-1} + 1$ (right). The dashed line (on right figure) shows the asymptotic limit for $\eta$ at $29.33\%$ when $N = 2^{n-1} + 1$. 
    Note that $n = \ceil{\log_2 (N)}$. 
    }
    \label{fig:etaplots}
\end{figure}

Fig.~\ref{fig:etaplots} illustrates the percentage improvement, $\eta$ for $M=1$ and arbitrary $N$ (left) and for $N = 2^{n-1} + 1$ (right).
In the case where $N = 2^{n-1} + 1$, depicted on the right, the plot demonstrates the specific improvements for this particular form of $N$. The dashed line indicates the asymptotic limit for $\eta$ at 29.33\%, illustrating that for $N$ of this form, the percentage improvement in oracle calls and Grover iterations stabilizes at 29.33\%.

\begin{remark}
 One can replace $\mathcal{A}$ with $U_N$ in the amplitude amplification algorithm and use it advantageously for several applications (in addition to Grover's search algorithm) that involve amplitude amplification. Such applications include quantum optimization~\cite{nielsen2002quantum,shukla2019trajectory}, quantum machine learning~\cite{rajagopal2021quantum}, cryptanalysis~\cite{montanaro2016quantum}, quantum chemistry~\cite{nielsen2002quantum} and genetic algorithms~\cite{acampora2022using}. 
\end{remark}

\subsection{Computational Example}

\begin{figure}[H]
    \centering
    \begin{minipage}[b]{0.45\textwidth}
        \centering
        \includegraphics[scale=0.5]{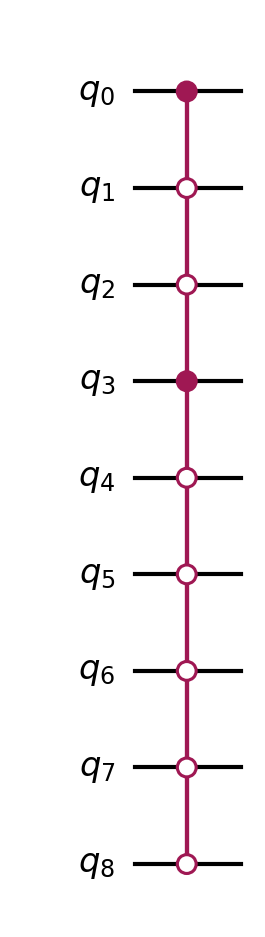}
        \caption{Oracle circuit block ($U_{\mathcal{P}}$) details for the case when  $\ket{9}$ is the only marked item out of a total of 273 items.}
        \label{fig:Uoracle}
    \end{minipage}
    \hfill
    \begin{minipage}[b]{0.45\textwidth}
        \centering
        \includegraphics[scale=0.5]{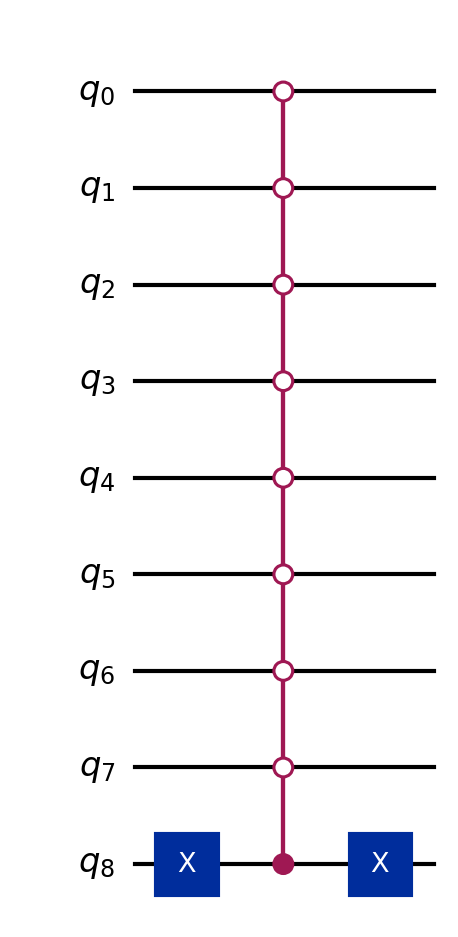}
        \caption{Circuit block ($U_0$) details for performing a phase flip for $\ket{0}$ state, for $N=273$ (with 9 qubits)}
        \label{fig:U0}
    \end{minipage}
    \end{figure}

\begin{figure}[H]
    \begin{minipage}[b]{0.9\textwidth}
        \centering
        \includegraphics[width=0.8 \textwidth]{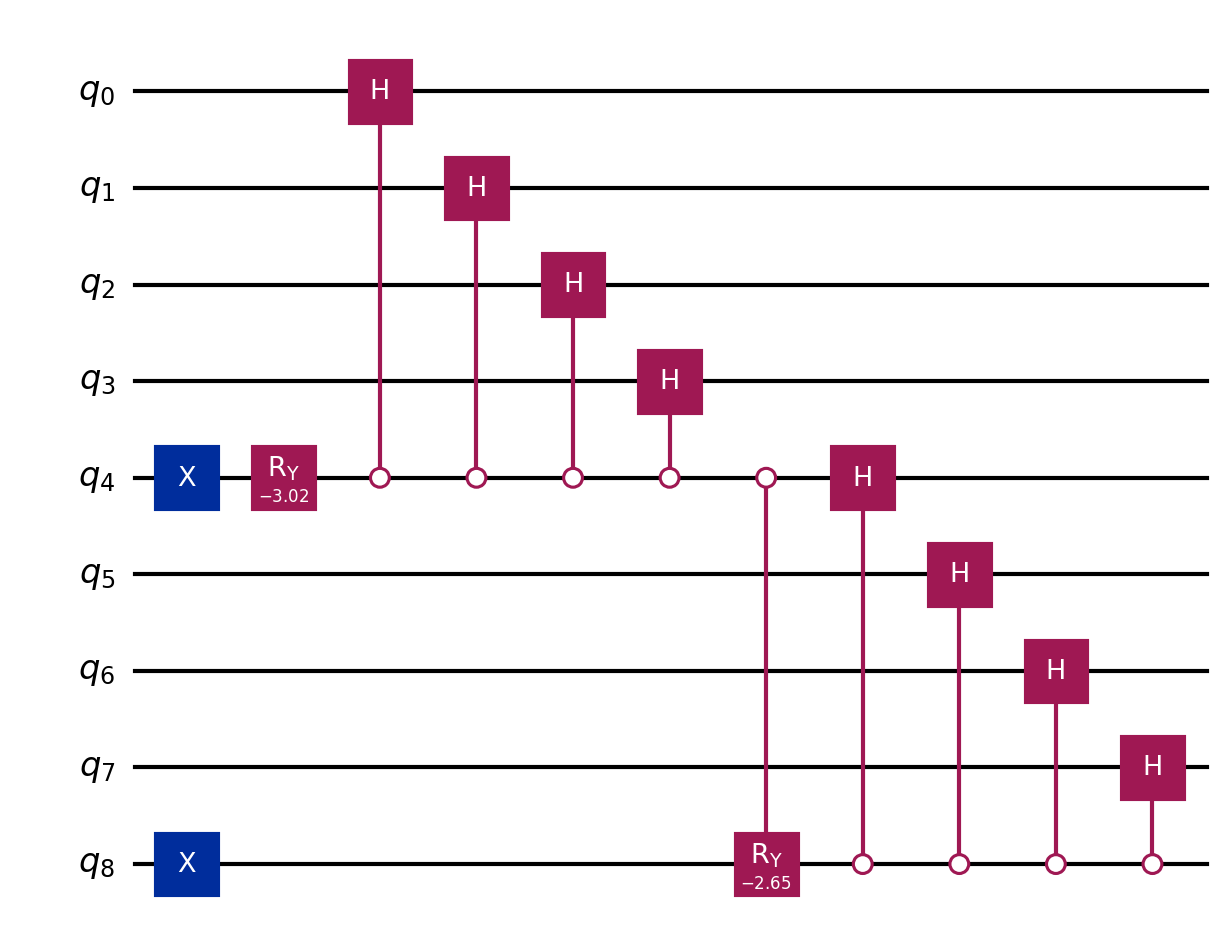}
        \caption{State preparation circuit block ($U_N$) details based on an efficient algorithm for preparation of uniform superposition states proposed in Ref.~\cite{shukla2024efficient}, for $N = 273$.}
        \label{fig:UN}
    \end{minipage}
\end{figure}

We consider a particular case of $N=273$ and $M=1$, with the computational basis state $\ket{9}$ being the marked item. 
In this case, the traditional Grover's algorithm requires $17$ oracle calls, whereas our proposed implementation requires only $12$ oracle calls for finding the marked item with probability close to $1$. 
This represents $29.41 \%$ reduction in the number of oracle calls compared to the traditional Grover's algorithm. 
The circuit diagrams for the implementation of unitary operators corresponding   $U_{\mathcal{P}}$, $U_0$ and $U_N$ are shown in Figures~\ref{fig:Uoracle}, \ref{fig:U0} and \ref{fig:UN}.
These circuits were created and executed within IBM's Qiskit simulation environment (version 1.0)~\cite{qiskit2024}. 
The results obtained were confirmed to be correct.

\section{Conclusion}
We present a novel and efficient implementation of a quantum search algorithm when the size of the search space $N$ is arbitrary and not limited to a power of $2$. While traditional implementation of Grover's search algorithm, involves creation of a uniform superposition state over $2^n$ computational basis states using Hadarmard gates over $n$ qubits, where $n=\ceil{\log_2 (N)}$, the proposed implementation uses an efficient method for creation of a uniform superposition state $U_N\ket{0}^{\otimes n} = \frac{1}{\sqrt{N}} \, \sum_{j=0}^{N-1} \ket{j}$  over $N$ computational basis states \cite{shukla2024efficient}. This implementation of state preparation has a gate complexity and circuit depth of $O(\log_2 (N))$ and does not require the use of any ancilla qubits. As a result, the proposed implementation requires $\left\lfloor \frac{\pi}{4} \sqrt{\frac{N}{M}} \right\rfloor$ oracle calls, which is considerably lower in comparison to the  number of oracle calls $\left(\left\lfloor \frac{\pi}{4} \sqrt{\frac{2^n}{M}} \right\rfloor\right)$ required in traditional implementation of Grover's search algorithm in many cases.

The improvements, relevant to reduced number of oracle calls and Grover iterations, are particularly notable when $N$ is slightly greater than a power of 2, with an asymptotic reduction of up to 29.33\% in oracle calls for such instances (for a fixed $M$, and as $N$ increases).


\end{document}